\documentstyle[12pt,dina4p]{article}
\title{\vskip-3cm{\baselineskip16pt
\centerline{\normalsize DESY 96--044\hfill ISSN 0418-9833}
\centerline{\normalsize hep-ph/9603385 \hfill}
\centerline{\normalsize March 1996\hfill}}
\vskip1.5cm
       \bf A New Calculation of the NLO Energy-Energy Correlation
           Function}
\author{
    { G.\ Kramer$^a$, H.\ Spiesberger$^b$} \\
        {$^a$II. Institut f\"ur Theoretische Physik${}^{*,\dagger}$} \\
        {Universit\"at Hamburg }\\{D - 22761 Hamburg, Germany}\\
        {$^b$Deutsches Elektonen-Synchrotron DESY}\\
        {D - 22603 Hamburg, Germany}\\}
\date{             }
\begin{document}
\maketitle
\vspace{2cm}
\begin{abstract}
We present a new calculation of the $O(\alpha_s^2)$ coefficient of the
energy-energy correlation function (EEC) using two different schemes
to cancel infrared and collinear poles. The numerical evaluation uses
the phase space slicing and the hybrid subtraction method. Both schemes
converge with decreasing slicing cut. The results are independent of the
scheme for small cuts. For the pure phase space slicing method, the cut
must be below $10^{-6}$ to achieve good results. All four approaches
agree with each other and confirm the results of Kunszt and Nason and
Glover and Sutton, for the latter also with respect to contributions of
different colour factors.
\end{abstract}
\vspace*{\fill}
 
\footnoterule
 
{\footnotesize
\noindent
${}^*$ Supported by Bundesministerium f\"ur Forschung und Technologie,
       Bonn, Germany, Contract 05\,6HH93P(5). \\
${}^{\dagger}$ Supported by the EEC Program 'Human Capital and
       Mobility' through Network 'Physics at High Energy Colliders'
       under Contract CHRX-CT93-0357 (DG12 COMA).
}
 
\newpage
 
\section{Introduction}
The energy-energy correlation function $\Sigma$ has been used by all
for LEP experiments \cite{Lep} and the SLD experiment \cite{SLD} at
SLAC to measure the strong coupling constant $\alpha_s$ in $e^+e^-$
annihilation at the Z resonance. $\Sigma$ is defined as a function of
the angle $\chi$ between two particles $i$ and $j$ in the following form
\begin{eqnarray}
\frac{d\Sigma(\chi)}{d\cos \chi} = \frac{\sigma}{\Delta \cos\chi
    N_{event}} \sum_{N_{event}} \sum_{i\neq j} \frac{E_iE_j}{E^2}
\end{eqnarray}
where $E_i$ and $E_j$ are the energies of the particles and $E$ is the
total energy of the event, $E^2=s$. The sum runs over all pairs $i,j$
with $\cos\chi$ in a bin of width $\Delta \cos \chi$: $\cos \chi -
\Delta\cos\chi/2 < \cos\chi < \cos\chi + \Delta\cos\chi/2$. Each
pair enters twice in the sum. The limits $\Delta\cos\chi \rightarrow 0$
and $N_{event} \rightarrow \infty$ have to be taken in (1). $\sigma$ is
the total cross section for $e^+e^- \rightarrow hadrons$.\\
 
In perturbative QCD the energy-energy correlation function (EEC)
is given as a series in $\alpha_s$ which we write as
\begin{eqnarray}
\frac{1}{\sigma_0} \frac{d\Sigma}{d\cos\chi}  =
\frac{\alpha_s(\mu)}{2\pi} A(\chi)
+\left(\frac{\alpha_s(\mu)}{2\pi}\right)^2
\left(\beta_0 \ln\left(\frac{\mu}{E}\right) A(\chi)
 +B(\chi)\right) + O(\alpha_s^3) ,
\end{eqnarray}
where $\beta_0 = (11N_c-4T_R)/3$. For QCD we have $C_F=4/3,~N_C=3$
and $T_R=N_f/2$ where $N_f$ is the number of active flavours at energy
$E$. \\
 
The first order term $A(\chi)$ is calculated from the well-known
one-gluon emission diagrams $\gamma^*,Z \rightarrow q\bar{q}g$, where
$\chi$ is the angle between any of the three partons. It has been first
calculated by Basham et al. \cite{Basham} with the result
\begin{eqnarray}
A(\chi)=C_F(1+\omega)^3\frac{1+3\omega}{4\omega}\left( (2-6\omega^2)
 \ln(1+1/\omega) +6\omega-3 \right)
\end{eqnarray}
where $\omega=\cot^2\chi/2$. The next-to-leading order (NLO)
contribution $B(\chi)$ is obtained from the processes $\gamma^*,Z
\rightarrow q\bar{q}g$ at one loop and $\gamma^*,Z \rightarrow
q\bar{q}gg,~q\bar{q}q\bar{q}$ at tree level. Several groups have
calculated $B(\chi)$ in the past: Richards, Stirling and Ellis (RSE)
\cite{RSE}, Ali and Barreiro (AB) \cite{AB}, Schneider, Kramer and
Schierholz (SKS) \cite{SKS}, Falck and Kramer (FK) \cite{FK}, Kunszt
and Nason (KN) \cite{KN}, Glover and Sutton (GS) \cite{GS} and just
recently Clay and Ellis (CE) \cite{CE}. Most of the calculations,
except those of SKS and FK, are based on the ERT matrix elements
\cite{ERT}. The numerical evaluations differ by the method used to
cancel infrared and collinear divergences.
 
Considering the results for $B(\chi)$ obtained by RSE, AB, FK and KN one
notices that they differ at the special point $\chi = \pi/2$ by roughly
$50 \%$. The SLD collaboration \cite{SLD} used all four theoretical
calculations to determine $\alpha_s$ from their measurements. They
averaged the values of $\alpha_s$ obtained this way and increased the
theoretical error accordingly. This additional error of $\alpha_s$ due
to the uncertainty over which the coefficient $B(\chi)$ in the EEC is
correct is equal to $\pm 0.006$, which is appreciable, considering that
the pure experimental error is only $(+0.002,-0.003)$ and the error from
the hadronisation corrections is $\pm 0.002$. This unsatisfactory
situation has raised renewed interest in the calculation of the EEC
function and Glover and Sutton \cite{GS} have presented a new
calculation of $B(\chi)$ using three methods, called subtraction, phase
space slicing and hybrid subtraction method. They confirmed the result
of Kunszt and Nason \cite{KN}. In the most recent calculation of Clay
and Ellis \cite{CE} a larger value of $B(\chi)$ is obtained which
agrees with the result of Falck and Kramer \cite{FK}, but is in
disagreement with the other recent calculations of GS and KN. CE
argued, on the basis of their much higher statistical precision of
$\pm 0.3\,\%$, that the theoretical error on $\alpha_s$ could be reduced
now; however, this does not seem justified since the spread of
theoretical predictions did not change with this new calculation. One
might think, that the theoretical prediction of the $O(\alpha_s^2)$
coefficient $B(\chi)$ should be unique with no other uncertainty than
the error of the numerical evaluation. However, we we will see later
that this need not always be the the case.\\
 
In order to compute the NLO term $B(\chi)$ in (2) it is necessary to
combine the contribution from the 3-parton one-loop diagrams
(multiplied with the LO graphs) with the 4-parton $\gamma^*,Z
\rightarrow q\bar{q}gg,q\bar{q}q\bar{q}$ processes. The virtual matrix
elements contain infrared and collinear singularities which cancel with
the singularities in the matrix elements of the 4-parton final state.
This cancellation of the infrared and collinear poles is done
analytically using dimensional regularisation. The extraction of the
infrared and collinear poles in the regulator $\epsilon$ (dimension
$n=4-2\epsilon$) and the complete calculation of the finite terms for
$\epsilon \rightarrow 0$ has been done by two groups independently with
identical results \cite{ERT,FKSS}. An analytical calculation of the full
4-parton expressions at finite $\epsilon$ is not possible. Therefore
these expressions are simplified in such a way that they contain all
the infrared and collinear singularities. Analytical integrals of these
simplified 4-parton matrix elements were then added to the virtual
3-parton contributions and the sum was shown to be finite in the limit
$\epsilon \rightarrow 0$. The difference between the exact 4-parton
matrix elements and the simplified expressions was then computed in 4
dimensions using numerical methods. To calculate this difference,
essentially two methods have been applied which are denoted subtraction
and phase space slicing method in the literature \cite{KS}. These two
methods can nicely be explained following Kunszt and Soper \cite{KS} by
considering a simple one-dimensional integral
\begin{eqnarray}
T = \lim_{\epsilon \rightarrow 0} \left(\int^1_0
    \frac{dx}{x}x^{\epsilon} F(x) - \frac{1}{\epsilon} F(0) \right)
\end{eqnarray}
where $F(x)$ is a known function representing the 4-parton matrix
elements. $x$ is the variable which produces the singularity, i.e.\ it
is the energy of the gluon, the angle of two partons or an invariant
mass which approaches zero in the infrared or collinear limit. The
integration over $x$ represents the additional phase space of the
parton which produces the pole term. The integrand is regularized by
the factor $x^{\epsilon}$ as it appears in dimensional regularization.
The first term is still divergent for $\epsilon \rightarrow 0$. So to
cancel the divergence of the second term, which represents the divergent
contribution from loop diagrams for the 3-parton matrix elements, one
must do the integral for $\epsilon \neq 0$. Now the two methods to
perform this integration are as follows. First in both methods one
isolates the singularity of the integrand by subtracting the pure pole
term $F(0)/x$ and adding it to the second term with the result
\begin{eqnarray}
T = \lim_{\epsilon \rightarrow 0}
    \left( \int_0^1 \frac{dx}{x} x^{\epsilon}
(F(x)-F(0)) + F(0) \int_0^1 \frac{dx}{x}x^{\epsilon}
  -\frac{1}{\epsilon}F(0)\right)
\nonumber    \\
  = \int_0^1 \frac{dx}{x} (F(x)-F(0))
\end{eqnarray}
This way the infrared singularity is cancelled and one is left with a
manifestly finite integral which is evaluated at $\epsilon = 0$. Due to
the complicated structure of the 4-parton matrix elements this last
integral must be calculated numerically. The above procedure defines
the subtraction method and was first used for the calculation of event
shape distributions in $e^+e^-$ annihilation by Ellis, Ross and Terrano
\cite{ERT}.\\

An alternative approach is the phase space slicing method which is
familiar from many QED calculations. It was first applied in a NLO
calculation of the thrust distribution in $e^+e^-$ annihilation by
Fabricius, Schmidt, Schierholz and Kramer \cite{FSSK}. In this method
the integration region in (4) is divided into two parts, $0<x<y_{min}$
and $y_{min}<x<1$. In the first region, the function $F(x)$ can be
approximated by $F(0)$ provided the arbitrary slicing parameter
$y_{min} \ll 1$, so that
\begin{eqnarray}
T = \lim_{\epsilon \rightarrow 0} \left( \int_{y_{min}}^1 \frac{dx}{x}
    x^{\epsilon} F(x) +
    \int_0^{y_{min}} \frac{dx}{x} x^{\epsilon} (F(x)-F(0))
    + F(0)\int_0^{y_{min}}\frac{dx}{x}
      x^{\epsilon} -\frac{1}{\epsilon} F(0)\right)
\nonumber  \\
= \int_{y_{min}}^1 \frac{dx}{x} F(x) +\int_0^{y_{min}}\frac{dx}{x}
  (F(x)-F(0)) +F(0)\ln y_{min}
\nonumber    \\
   \simeq \int_{y_{min}}^1 \frac{dx}{x} F(x) + F(0) \ln y_{min} .
\end{eqnarray}
The last equation in (67) follows when $y_{min} \ll 1$ is chosen small
enough, so that the finite integral
\begin{eqnarray}
 T_f = \int^{y_{min}}_0 \frac{dx}{x}\left(F(x) - F(0)\right)
\end{eqnarray}
can be neglected. If this is the case the integral $T$ should not
depend on $y_{min}$ which usually is taken as evidence that $T_f$ is
indeed negligible. This cutoff dependence was investigated, for
example, for the thrust distribution in \cite{GKS} and for $y_{min}$
small enough ($y_{min} \leq 10^{-4}$, where $y_{min}$ was the
invariant mass cutoff) the result was found to agree with the thrust
distribution obtained with the subtraction method \cite{ER,K}. It is
clear that if the finite contribution $T_f$ in (7) is not neglected both
the subtraction and the phase space slicing method must give the same
results, independent of how small the cutoff $y_{min}$ is. If the finite
term is kept, we have for $T$
\begin{eqnarray}
 T = \int^1_{y_{min}} \frac{dx}{x}F(x)  + F(0)\ln y_{min}  +T_f .
\end{eqnarray}
This version is called the hybrid subtraction method in the recent work
of Glover and Sutton \cite{GS} and will be called the hybrid method in
the following.\\
 
It is clear that in this simple example, the results for $T$ in the
subtraction method and in the hybrid method are identical, independent
of how $y_{min}$ has been chosen in (8). However, in the actual
application to the EEC the situation is more complicated. In this case
the function "$F(x)/x$" is the matrix element of the process $e^+e^-
\rightarrow 4~ partons$ which depends on five variables, whereas
"$F(0)$" corresponds to a contribution to the 3-parton final state,
$e^+e^- \rightarrow 3~partons$, which depends only on two variables.
Both contributions have to be integrated out up to the one variable
$\cos \chi$ and in order to complete the definition of the EEC function
one has to supply a prescription for the calculation of $\cos \chi$.
In the region $x \geq y_{min}$ where $x$ stands for one of the variables
for which the matrix elements become singular, it is most natural to
determine $\cos \chi$ from the pairing of two of the 4 parton momenta
$p_1,p_2,p_3$ and $p_4$, i.e.\
\begin{eqnarray}
 \cos \chi = \hat{\vec{p}}_i \hat{\vec{p}}_j
\end{eqnarray}
with $i,j =1,...,4$. The finite integral $T_f$ in (7) can be calculated
at least in two ways: (i) the integrand "$F(x)/x$" is calculated with
4-parton kinematics, i.e.\ with (9) and $i,j = 1,...,4$, whereas the
subtracted integrand "$F(0)/x$" is calculated with variables
corresponding to 3 partons in the final state, i.e.\ with (9) and
$i,j=1,..,3$, of course after the integration over $x$ and two more
variables is performed in the 4-parton case; (ii) the whole integral
with integrand $(F(x)-F(0))/x$ is evaluated with 3-parton kinematics.
The latter procedure amounts to performing the integration over $x$ and
two more of the 4-parton variables with two variables considered
constant which are then identified with the two variables describing a
3-parton final state. With other words, for the evaluation of $T_f$ one
has the freedom to treat the contribution $F(x)/x$ exactly or with
recombination of two of the partons into one jet. The selection of the
partons $i,j$, which are recombined is determined by that one of the
variables $y_{ij}=(p_i+p_j)^2/s$ which leads to the singularity and
which is identified with $x$ in the example integral. It is clear that
this recombination procedure is not unique and we shall consider two
possibilities later on. In the limit $y_{min} \rightarrow 0$ the
distinction (i) and (ii) should not matter. But for finite $y_{min} >
0$ we expect a difference between (i) and (ii). The procedure (i) is
used in the subtraction approach, so that for (i) the result of the
hybrid method should be independent of $y_{min}$ whereas with (ii) the
region where the recombination is performed is changed so that we
expect that the total integral will depend on $y_{min}$. The $y_{min}$
dependence should be smaller than in the phase space slicing approach,
in which the integral $T_f$ is neglected altogether. It is the general
consensus that all these methods should give legitimate results,
provided $y_{min}$ is chosen small enough.\\
 
As in previous work we focus on the zero-resolution limit ($y_{min}
\rightarrow 0$), i.e.\ we do not apply a recombination procedure to all
of the two-parton combinations to form jets before the calculation of
the EEC function is done. Calculations with a jet recombination for all
possible parton pairings were performed by AB \cite{AB} and by SKS
\cite{SKS} using Sterman-Weinberg ($\epsilon,\delta$) cuts and also in
a recent experimental analysis of the ALEPH collaboration
\cite{Aleph}.\\
 
To shed some further light on the question where the discrepancy
between the various calculations could arise we have made the effort to
perform a new calculation of $B(\chi)$ using the hybrid method in
the form (ii) described above. In recent work \cite{KSp} we have
studied the asymptotic behaviour of $B(\chi)$ for $\chi \rightarrow
\pi$ in order to compare with the predictions of the leading logarithm
approximation for large angles \cite{Log} using the hybrid method in
the same form. Here we extend these calculations to the whole $\chi$
range. In addition, we separated the contributions of the different
phase space regions, which can be characterized by the first, second and
third term in (8) for the example integral. Furthermore to explore the
source of the discrepancies we decompose $B(\chi)$ into the
contributions from different colour factors
\begin{eqnarray}
 B(\chi) = C_F^2 B_{C_F}(\chi) +C_FN_CB_{N_C}(\chi) +C_FN_fB_{N_f}(\chi)
\end{eqnarray}
which can be compared with the results of other calculations. In
section 2 we describe the two methods used to calculate the
$O(\alpha_s^2)$ contribution to the EEC and present the results for the
various colour factors separated according to the different phase space
regions. In section 3 we compare our results with the earlier
calculations and comment on the different approaches. We end with some
concluding remarks.\\
 
 
\section{Calculational Methods and Results}
The new calculation of the EEC in $O(\alpha_s^2)$ is based on the
approach described in detail in \cite{KL}. It starts from known
$O(\alpha_s^2)$ matrix elements for $e^+e^- \rightarrow 3\,partons$
\cite{ERT,FKSS} and $e^+e^- \rightarrow 4\,partons$ \cite{Ali,ERT}. To
obtain finite results, in which all infrared and collinear
singularities are cancelled we introduce a phase space slicing cut to
separate the 4-parton phase space from the region in which the
integration over one of the invariants $y_{ij}$ producing the
singularity has been done analytically in $n$ dimensions. This
integration region is defined by
\begin{eqnarray}
    y_{ij} = (p_i+p_j)^2/s < y_{min}
\end{eqnarray}
where $y_{min}$ is the parameter to separate the two regions and $i,j$
are the labels for two of the 4 parton momenta $p_i, i=1,2,3,4$. This
slicing procedure with an invariant mass cut is most convenient for
analytical calculations. The integration over only one of the
invariants is possible only when the singular contributions have been
separated. To achieve this a partial fraction decomposition of the
4-parton matrix elements is carried out (see \cite{KL} for details).
Then the EEC function is calculated from three separate contributions:\\
(a) The first contribution contains the singular parts of the partial
fractioned 4-parton matrix elements integrated over one invariant
inside the slicing cut $y_{min}$ together with the virtual corrections
to the $q\bar{q}g$ final state. This contribution corresponds to the
term $F(0)\ln y_{min}$ in (8) for the example integral and depends
strongly on $y_{min}$ with the dominant term $\propto
(-\ln^2y_{min})$.\\
(b) The second contribution contains the remaining non-singular part of
the 4-parton matrix elements which is the difference between the full
4-parton matrix element and the singular part already included in (a)
integrated over the same region as in (a). This part corresponds
obviously to the term $T_f$ in (8). \\
(c) The third part consists of all contributions outside the singular
region (a), i.e.\ with the intgration over $y_{ij}>y_{min}$, which is
computed numerically with $\cos \chi$ given by (9) with $i,j=1,2,3,4$.
This part represents the first term in the example integral on the
right-hand side of (8). \\
The calculation of the 4-parton contributions needed in (a) proceeds as
follows. First the cross section for $e^+e^- \rightarrow q\bar{q}gg$
(the $q\bar{q}q\bar{q}$ final state is less singular and does not
require partial fractioning; otherwise it is treated similarly) has the
general form
\begin{eqnarray}
 d^5\sigma = \left(\frac{\alpha_s}{2\pi}\right)^2 f(y_{ij})dPS^{(4)} .
\end{eqnarray}
The right-hand side of (12) has pole terms proportional to $y_{ij}^{-1},
(ij=13,14,23,24,34)$ which are separated by the partial fractioning.
For example, the contribution proportional to the colour factor $C_F^2$
has the structure
\begin{eqnarray}
 f(y_{ij}) = \frac{A_{13}}{y_{13}} +  (1 \leftrightarrow 2) +
 (3 \leftrightarrow 4) + (1 \leftrightarrow 2, 3 \leftrightarrow 4)
\end{eqnarray}
where the momentum labels are $1,2,3,4=q,\bar{q},g,g$. The terms
proportional to $y_{13}^{-1}$ and $y_{14}^{-1}$, respectively
$y_{23}^{-1}$ and $y_{24}^{-1}$, become singular when one of the gluons
is infrared or collinear with the quark, respectively antiquark. They
produce the dominant negative contributions to $B(\chi)$ after
integration over the unresolved regions $y_{13}<y_{min}$,
$y_{23}<y_{min}$, etc.\ when they are added to the virtual
contributions. This integration is done only for one of the four terms
in (13) which are related by permutation of the momentum labels. It is
important to note that with this procedure of slicing we do not
calculate genuine 3-jet cross sections since only one of the pairings
$ij$ of two partons is recombined into one jet in the first, second
etc. term in (13). The four terms in (13) are separated into singular
and non-singular terms. The singular terms are regularized by
dimensional regularization. The singularities in $\epsilon$ after
integration compensate against the singularities in the $O(\alpha_s^2)$
one-loop corrections to $e^+e^- \rightarrow q\bar{q}g$. The result from
the contribution (a) will be denoted the singular contribution,
abbreviated by {\tt sing} in the figures and tables which contain our
results.\\
 
The other two contributions (b) and (c) come exclusively from the
4-parton tree diagrams. The part (b) consists of the non-singular
terms in $A_{13}/y_{13}$ etc. in (13) which are also integrated up to
the cutoff. They do not participate in the cancellation of the
infrared/collinear singularities between tree and virtual diagrams. The
third part (c) is connected with the contribution above the slicing
cut, $y_{13}>y_{min}$ (in the first term of (13)), which is computed
numerically. It is important to note that the same expression for
$f(y_{ij})$ with partial fractioning is used as in the singular region
and that the integration up to $y_{min}$ over $y_{13}$ is applied only
to the first term in (13). In the other terms the integration is over
the variables $y_{23}, y_{14}$ and $y_{24}$. The partial fractioning is
not unique in the sense that non-singular terms can be distributed
between the terms proportional to $A_{ij}$ in (13). In the figures and
tables and in the discussion below the contributions (b) and (c) will
be called finite terms ({\tt fin}) and real terms ({\tt real}),
respectively.\\
 
To proceed with the calculation we must specify the singular
contributions given by $A_{ij}/y_{ij}$ in (13). These singular terms
factorize into the pole terms $y_{ij}^{-1}$ times a factor which can be
identified with the LO matrix element for $e^+e^- \rightarrow
q\bar{q}g$. Only because of this factorization one is able to perform
the cancellation with the virtual contributions. The identification of
this factor with the LO matrix element amounts to specifying
combinations of the 4-parton kinematic variables for the $q\bar{q}gg$
final state which define a 3-jet final state. These 3 jets can then
be identified with the 3 partons of the LO matrix element. Usually this
is done with the help of invariants built from the momenta of the
$q\bar{q}gg$ final state. For example in the ERT approach \cite{ERT}
which was followed in all the calculations using the subtraction
method, the relation was such, that in the pole term proportional to
$y_{ij}^{-1}$ the 3-jet variables are $y_{134},y_{24}$ and $y_{123}$
($y_{ijk}=(p_i+p_j+p_k)^2/s$). For three massless jets one has
$y_{134}+y_{24}+y_{123}=1$. This agrees with the 4-parton
energy-momentum relation $y_{134}+y_{24}+y_{123}-y_{13}=1$ only in the
limit $y_{13} \rightarrow 0$. Therefore in \cite{KL} two other schemes
have been considered: (1) the so-called KL scheme where
$y_{134},y_{24}$ and $y_{123}-y_{13}$ are used instead; (2) the
so-called KL' scheme, where the 3-jet variables are
$y_{134},y_{24}-y_{13}$ and $y_{123}$. Of course there are many more
possibilities. This non-uniqueness of the definition of 3-jet variables
for a 4-parton final state can not be avoided if one wants to cancel
the infrared and collinear singularities between real and virtual
contributions before the distribution in the final observable, i.e.,
in our case, in $\cos \chi$, is computed. This difference in the 3-jet
variables is supposed to have no effect for $y_{min} \rightarrow 0$,
i.e.\ in this limit we expect no difference in the results for the KL,
KL' and ERT scheme. Since $y_{min}$ is really never put equal to zero
we must anticipate an effect from these different schemes in practice.
From this discussion it is clear that in the calculation of the EEC
function there is an ambiguity which is not visible in the simplified
example integral discussed in the introduction. There, $F(0)$ can be
defined uniquely: since $F$ depends only on one variable, $F(0)$ is the
residue of the pole. With more than one variable, however, the
equivalent of $F(0)$ depends on which of the other variables are kept
fixed and identified as 3-jet variables. \\
 
In the following we present separate results for the contributions of
the $C_F^2,C_FN_C$ and $C_FN_f$ parts  of $B(\chi)$ as defined in (10),
where we include the colour factor, i.e.\ we plot
$\sin^2\chi\;C_F^2 B_{C_F}(\chi)$ etc.\ for the three contributions
{\tt sing}, {\tt fin} and {\tt real} for a slicing cut
$y_{min}=10^{-4}$. We have chosen the KL' scheme for these plots and
factored out $\sin^2\chi$ to be able to plot with a linear scale. In
Fig.\ 1 we have plotted the contributions for the $C_F^2$ part as a
function of $\cos \chi$. We see that the real part is large and
positive whereas the singular part is large and negative as we expect.
The sum of both terms is small and negative for $\cos \chi \rightarrow
\pm 1$, since the singular terms $\propto \ln^n y_{min},n=1,2$, cancel
in the sum. The finite term ({\tt fin}) is very small. In the figure it
is hard to distinguish it from the zero line. The same plot for the
$C_FN_C$ part is seen in Fig.\ 2. Here we have two classes of pole
terms coming either from the $qg(\bar{q}g)$ or the $gg$ recombination.
The corresponding real  and non-singular parts are denoted {\tt
real(13)}, {\tt fin(13)} and {\tt real(34)}, {\tt fin(34)},
respectively. In the singular term both contributions (13) and (34) are
added together for convenience. We see that the {\tt fin(34)} is larger
than {\tt fin(13)} and {\tt real(34)} ({\tt real(13)}) is large
positive (large negative). The total sum is positive. For the $C_FN_f$
contribution, calculated for $N_f=5$ and exhibited in Fig.\ 3, we have
only the singular and the real contributions. The finite terms are
included already in the singular terms (see \cite{KL} for details). The
singular and the real contributions contain terms which are
proportional to $(-\ln y_{min})$ and $\ln y_{min}$, respectively. These
terms cancel in the sum which is negative and smaller. In addition we
have plotted the contribution of the $q\bar{q}q\bar{q}$ interference
term. This term is actually proportional to $C_F(C_F-N_C/2)$ and should
be distributed to the $C_F^2$ and $C_FN_C$ terms. For convenience we
have included it here. Since it has no singularities it can be
calculated numerically without difficulties. Its contribution is very
small and negligible compared to the other terms. The sum in Fig.\ 3
includes this interference term, i.e.\ the sum contains all 4-quark
contributions. In Fig.\ 4 the $C_F^2,C_FN_C$ and the $C_FN_f$
contributions are plotted together with the sum of all three colour
terms. We see that the $C_FN_C$ term is the most important one and it
is positive. The $C_FN_f$ term is smaller and negative. The $C_F^2$
term is less important. The sum is essentially given by the sum of the
$C_FN_C$ and the $C_FN_f$ parts except near the wings of the EEC
distribution. The corresponding curves for the KL scheme look similar
and will not be shown here. For larger $y_{min}$ values the  results
change. For the KL' scheme the EEC correlation function decreases with
increasing $y_{min}$. For the KL scheme this decrease occurs in
particular for the wings of the NLO EEC function. \\
 
To obtain an overview of the most interesting central region we give
the numbers in Tab.\ 1 for $\cos \chi = 0$ and $y_{min}=10^{-k}$,
$k=2,3,4,5,6$. Since the numerical results fluctuate from bin to bin we
have made a polynomial fit to our results in the interval $-0.40<\cos
\chi <0.40$ and quote the result of this fit at $\cos \chi =0$. We give
the results for the different colour factors. We see that the $C_FN_f$
term is independent of $y_{min}$, which we expect because of the less
singular behaviour of this part. The variation of the EEC function
results in particular from the variation of the $C_F^2$ term, whereas
the variation of the $C_FN_C$ contribution with $y_{min}$ is much
smaller. In the KL case the variation of the $C_F^2$ contribution is
smaller, so that also the total sum is fairly constant. The splitting
up of $B(\chi )$ into the contributions from the various regions for
$\cos \chi = 0$ and $y_{min}=10^{-4}$ is given in Tab.\ 2 for the KL'
case. We see that the finite part {\tt fin(34)} is still not negligible
for this slicing cut. It is still more than $10 \%$ of the total sum
and larger than the absolute value of the total $C_F^2$ contribution.
The dependence of this particular contribution on the value of the
slicing cut is shown in Fig.\ 5. The $q\bar{q}q\bar{q}$ interference
term is negligible and much smaller than the total error. \\
 
Results for the pure phase space slicing method can be obtained by
subtracting the contributions of the finite terms from the results in
Tab.\ 1. The resulting numbers for the same $y_{min}$ as in Tab.\ 1 are
collected in Tab.\ 3, again for the different colour factors and the KL'
and KL schemes. We see that the convergence with $y_{min}$ is slower
than for the hybrid method. The slow convergence comes mostly from
neglecting the term {\tt fin(34)} which is the largest one of all the
finite contributions (see Tab.\ 2). Even for the larger $y_{min}$
values the results for the KL' and KL approaches differ very little.
The difference between the two schemes in Tab.\ 1 comes essentially
from the finite term in the $C_F^2$ contribution which changes stronger
with $y_{min}$ in the KL' scheme. The slower convergence of the pure
phase space slicing method was also observed by Glover and Sutton
\cite{GS}. If we compare the results for the hybrid and the phase space
slicing method at $y_{min}=10^{-6}$ we see that they agree inside the
numerical errors. The difference of $B(\chi=\pi/2)$, however, is still
$1.15$. This comes essentially from the term {\tt fin(34)} which
decreases very slowly with decreasing $y_{min}$, but is still
non-negligible. In the calculation we obtained ${\tt fin(34)}=1.154 \pm
0.006$ for both the KL' and the KL case. So, even with a slicing cut
as small as $10^{-6}$ the finite parts are still not very small and
produce an error of approximately $2\%$ for $B(\chi =\pi/2)$ if one
uses the pure phase space slicing method. At $y_{min}=10^{-5}$ the
$B(\chi)$ differ already outside the given error. Thus, to obtain good
values for the NLO coefficient $B(\chi)$ with the phase space slicing
method one must go to extremely small $y_{min}$ of the order of
$10^{-6}$ or smaller. The dependence of $B(\chi=\pi/2)$ on $y_{min}$ is
shown in Fig.\ 6 for the hybrid and space space slicing method and for
the KL' and KL scheme separately for $y_{min}$ between $10^{-6}$ and
$10^{-2}$. We observe that there is little difference between the KL'
and the KL scheme in the case of the phase space slicing procedure
since these two schemes differ essentially only in the finite parts.
For the phase space slicing method, $B(\chi=\pi/2)$ increases
monotonically with decreasing $y_{min}$. With the hybrid method
$B(\chi)$ first increases to a maximum near $y_{min}=10^{-3}$ and then
decreases towards the final value at $y_{min}=10^{-6}$. For $y_{min}
\leq 10^{-4}$ the two schemes, KL' and KL, give the same results inside
the errors. We consider the value at $y_{min}=10^{-6}$ which is
$B(\chi=\pi/2)=50.1 \pm 0.9$ as our final value. In Fig.\ 7 we have
plotted $B(\chi)$ as a function of $\chi$ in the range $-0.96 < \cos
\chi < 0.96$ for the same $y_{min}$ for KL', as well as the result for
$y_{min} = 10^{-4}$.\\
 
 
\section{Comparison with other Results}
In this section we compare with the results of previous calculations.
Such a comparison was done already by Glover and Sutton who compared
their results with those of AB, RSE, FK and KN. Therefore we can
restrict ourselves to a comparison with the most recent evaluations,
namely those of KN, GS and CE. It turns out that our results agree with
those of KN and GS but not with the results of CE. Kunszt and Nason
\cite{KN} calculated $B(\chi)$ with the subtraction method by
reorganizing the ERT matrix elements to give numerically stable
results. Glover and Sutton \cite{GS} determined $B(\chi)$ with all three
methods, subtraction, phase space slicing and hybrid method. For the
evaluation with the subtraction method they used the ERT matrix
elements as given in the original work \cite{ERT}. For the phase space
slicing and the hybrid method they have constructed a completely
independent program based on \cite{GG} but using squared matrix
elements rather than helicity amplitudes as in \cite{GG}. Our
calculation is based on the work reported in \cite{KL} which was
developed independently of all the earlier results and which relied on
the introduction of a slicing cut to isolate the infrared and
collinear divergences. A further important ingredient was the use of
partial fractioning in all invariants $y_{ij}$ which lead to
singularities so that only for one of these variables a slicing cut had
to be introduced. We believe that a similar technique was employed by
Kunszt and Nason and by Glover and Sutton in their calculations. \\
 
In Fig.\ 6 we can see that at larger $y_{min}$ the results vary with
$y_{min}$, independently of the method that was used; even for $y_{min}
< 10^{-5}$ we still have a non-negligible variation. Since on the other
hand  Glover and Sutton give explicit numerical results only for
$y_{min}=10^{-5}$ we have selected our results for the same $y_{min}$
to perform the comparison. Since our calculational methods and
those of Glover and Sutton are not the same, there is actually no reason
that the two results should be identical for this particular $y_{min}$;
however, we believe that a comparison at the same $y_{min}$ is more
sensible than performing the comparison at different values. Since
Kunszt and Nason used the subtraction method, there is no cut
dependence in their results.\\
 
The comparison of our results with the results of these two groups is
collected in Table 4. We see that we have perfect agreement. In
particular our hybrid results agree very well with those of Glover and
Sutton inside their respective errors (compare $KS^{(5)}, KS^{(6)}$
with $GS^{(4)}$). Concerning the phase space slicing method there is a
small difference with GS outside the errors (compare $KS^{(7)},
KS^{(8)}$ with $GS^{(3)}$). Knowing that the finite parts are still
non-negligible at $y_{min}=10^{-5}$, we can not expect perfect
agreement in this case.\\
 
Another important check between the various results is the
decomposition of $B(\chi)$ into contributions for different colour
factors. In Tables 1 and 3 we have given these contributions already for
$B(\chi)$ at $\chi=\pi/2$. Equivalent results are not available from
the publications in refs.\ \cite{KN,GS}. After their publication,
Glover and Sutton calculated the contributions for the different colour
factors for a comparison with the results of Clay and Ellis. Their
results for $B(\chi=\pi/2)$ \cite{G} are:
\begin{eqnarray}
  C_F^2B_{C_F}=-2.42 \pm 0.92,\\
\nonumber
  C_FN_CB_{N_C}=77.20 \pm 2.08,\\
\nonumber
  C_FN_fB_{N_f}=-24.97 \pm 0.53.
\end{eqnarray}
The sum of these terms is $49.8 \pm 2.3$. By comparing with our results
in Table 1 or Table 3 we observe perfect agreement of our colour
decomposition with the Glover and Sutton results given above. \\
 
Concerning the more recent results of Clay and Ellis \cite{CE} there is
no agreement with our results and with those of Kunszt-Nason or
Glover-Sutton, respectively. From Fig.\ 1 of the Clay-Ellis paper we
read off a value of about $57.0$ for $B(\chi=\pi/2)$. The actual
precision of their numerical results is $0.3\%$ as stated by the
authors. So, our value of $B$ as well as that of KN or of GS differ by
approximately $15\%$ which can not be explained by numerical
uncertainties in any one of the calculations. In our case the
error is below $2\%$ (for $y_{min}=10^{-6}$). The difference is in the
contributions with the colour factors $C_F^2$ and $C_FN_C$, since
concerning $B_{N_f}$ Clay and Ellis and Glover and Sutton achieved
agreement \cite{CE}. It is clear that the $C_FN_f$ term is less
problematic since the contributions to this term are less singular than
those to the $C_F^2$ and $C_FN_C$ terms. The origin of the
disagreement is unclear. Unfortunately we could not study this further,
since Clay and Ellis have not explained the details of their
calculation.\\
 
With respect to the earlier calculations of AB \cite{AB}, RSE
\cite{RSE} and FK \cite{FK} we made some effort to understand the
differences with these earlier results. The calculations of FK were done
with the phase space slicing method with $y_{min}=10^{-4}$ and with the
important difference that the real contributions are calculated with
cuts $y_{ij} \geq 10^{-4}$ applied to all invariants of the 4-parton
momenta. This was necessary since for the real contributions partial
fractioning as in (13) was not performed. We found out, however, that
for $y_{min}=10^{-4}$ these additional cuts have a non-negligible effect
when we apply them to the partial fractioned real term as given, for
example, in (13). In the present calculation we used cuts only for
those variables which are singular in the first, second etc. terms in
(13). In \cite{FK} the reduction of the non-singular terms through
these additional cuts was compensated by additional terms in the
singular contributions, but presumably not fully. In the moment it
is unclear, whether with much smaller cut values than $10^{-4}$ the
method used by FK would give the same results as the present
calculation. From the results given by FK it is clear that
$y_{min}=10^{-4}$ is not sufficiently small to obtain correct results.
Furthermore, from Tables 1 and 3 we see that at $y_{min}=10^{-4}$ there
is already a difference of $\Delta B(\chi =\pi/2) =6.03$ and $6.45$
for the KL' and KL schemes, respectively, between the hybrid and the
phase space slicing method. The results of AB \cite{AB} and RSE
\cite{RSE} are obtained with the subtraction method and therefore
should agree with the results of KN and GS using the same method. We
suspect that in these older calculations the so-called finite pieces
(see (5)) are calculated also with cuts on all invariants $y_{ij}$ in
the 4-parton term (first term in (5)). In the work of Ali and Barreiro
(first reference of \cite{AB}) such a cut $y_{ij} \geq 10^{-7}$ is
explicitly stated and this cut was also used in the second paper
\cite{AB,A}. Concerning RSE we have no information on this point.
If the interpretation that the additional cuts in AB and RSE are
responsible for the reduction of their results is correct, it would
mean that, similar as we observed for the FK calculation, one would
need even smaller cuts below $10^{-7}$ to achieve convergence.\\
 
We also compared the EEC coefficient $B(\chi)$ for the other angles as
shown in Fig.\ 7 with the results of Kunszt and Nason and found good
agreement. We therefore conclude that the KN calculation gives the
correct NLO coefficient $B(\chi)$. The older results RSE, AB and FK
should be disregarded and not be used for a determination of $\alpha_s$.
If for the time being we also leave aside the results of Clay and Ellis
\cite{CE} since their result has not been confirmed by other
calculations, the theoretical error on $\alpha_s$ mentioned in the
introduction can be eliminated. Then the SLD analysis \cite{SLD} yields
the following value for $\alpha_s$ from the measurement of the EEC:
\begin{equation}
\alpha_s(M_Z^2) = 0.125_{-0.003}^{+0.002}\,({\rm exp})
                  \pm {}0.012 \,({\rm theory}).
\end{equation}
This agrees completely with the conclusion of Glover and Sutton
\cite{GS}. Now the theoretical error comes exclusively from the scale
change which in the SLD work \cite{SLD} originates from a rather
large variation with a scale factor $f$ in the interval $0.002 \leq
f \leq 4$.\\
 
In conclusion, motivated by the recent results of Clay and Ellis which
disagreed with the earlier results of KN and GS, we have
reevaluated the EEC using a completely independent approach which
allowed us to obtain results for the phase space slicing and the
hybrid method. The pure phase space slicing method needs extremely
small slicing cuts, smaller than $10^{-6}$, in order to achieve a good
accuracy below $2\,\%$. This is due to the finite terms which converge
very slowly to zero with decreasing cut. This shows that configurations
with extremely small invariant masses can contribute to the EEC at all
angles, i.e.\ also at angles far away from the two-jet region at $\chi
\simeq \pi$. In our opinion the results of the older calculations
seem to suffer from a too large slicing cut (FK) or from additional
cuts in the evaluation using the subtraction method (RSE, AB) and
therefore could yield only estimates of $B(\chi)$.\\
 
\vspace{5mm}
\noindent
{\Large \bf Acknowledgements} \\
 
\noindent
We thank E.\ Glover for communicating his results on the colour
decomposition and A.\ Ali for useful discussions on his work with
F.\ Barreiro. \\
 
\vspace{1cm}
 
\noindent
{\Large \bf Table Caption}\\
\begin{list}{}{\setlength{\leftmargin}{18mm}%
\setlength{\labelwidth}{15mm}}
\item[Tab.\,1: \hfill]
      NLO coefficient $B(\chi)$ at $\cos \chi = 0$ for various $y_{min}$
      and colour contributions in the KL' and KL scheme using the
      hybrid method.
\item[Tab.\,2: \hfill]
      Contribution to the NLO coefficient $B(\chi)$ at $\cos \chi = 0$
      for $y_{min} = 10^{-4}$ for different colour factors from the
      regions {\tt sing}, {\tt real} and {\tt fin} with the hybrid
      method and KL' scheme.
\item[Tab.\,3: \hfill]
      Same as in Table 1 with the phase space slicing method.
\item[Tab.\,4: \hfill]
      NLO coefficient $B(\chi)$ at $\cos \chi =0$ for different
      calculations; (1) Table 3 of \cite{KN}, (2) subtraction method,
      (3) phase space slicing method, $y_{min}=10^{-5}$, (4) hybrid
      method, $y_{min}=10^{-5}$ of ref. \cite{GS}, (5) and (6) phase
      space slicing method for KL' and KL scheme from Table 3, (7) and
      (8) hybrid method for KL' and KL scheme from Table 1
      ($y_{min}=10^{-5}$).
\end{list}
 
\vspace{1cm}
 
\noindent
{\Large \bf Figure Caption}\\
\begin{list}{}{\setlength{\leftmargin}{18mm}%
\setlength{\labelwidth}{15mm}}
\item[Fig.\,1: \hfill]
      $C_F^2$ contributions to the NLO coefficient $\sin^2\chi\;
      B(\chi)$ as a function of $\cos \chi$ for $y_{min} = 10^{-4}$ in
      the KL' scheme. Long-dashed line: singular contribution,
      short-dashed line: real contribution, dotted line: finite
      contribution. Full line: sum of all $C_F^2$ contributions.
\item[Fig.\,2: \hfill]
      $C_FN_C$ contributions to the NLO coefficient $\sin^2\chi\;
      B(\chi)$ as a function of $\cos \chi$ for $y_{min} = 10^{-4}$ in
      the KL' scheme. Long-dashed line: singular contribution; real
      contributions: short-dashed: {\tt real(13)}, dotted:
      {\tt real(34)}; finite contributions: long dashed-dotted: {\tt
      fin(13)}, short dashed-dotted: {\tt fin(34)};
      full line: sum of all $C_FN_C$ contributions.
\item[Fig.\,3: \hfill]
      4-quark contributions to the NLO coefficient $\sin^2\chi\;
      B(\chi)$ as a function of $\cos \chi$ for $y_{min} = 10^{-4}$ in
      the KL' scheme. Long-dashed line: singular contribution, dotted
      line: real contribution; short-dashed line: 4-quark interference
      contribution. Full line: sum of all 4-quark contributions.
\item[Fig.\,4: \hfill]
      Colour decomposition of the NLO coefficient $\sin^2\chi\;
      B(\chi)$ as a function of $\cos \chi$ for $y_{min} = 10^{-4}$ in
      the KL' scheme. Long-dashed line: $C_F^2$ contribution,
      short-dashed line: $C_FN_C$ contribution, dotted line: $C_FN_f$
      contribution; full line: sum of all contributions.
\item[Fig.\,5: \hfill]
      $y_{min}$ dependence of $C_FN_C$-{\tt fin(34)} contribution to
      the NLO coefficient $B(\chi=\pi/2)$.
\item[Fig.\,6: \hfill]
      $y_{min}$ dependence of the NLO coefficient $B(\chi=\pi/2)$.
      Upper curves are for the hybrid method, lower curves for the pure
      phase space slicing method. The two curves in each set are for
      the KL' and the KL scheme.
\item[Fig.\,7: \hfill]
      The complete NLO coefficient $B(\chi)$ as a function of $\cos
      \chi$ obtained with the hybrid method in the KL' scheme for
      $y_{min} = 10^{-4}$ (histogram) and $y_{min} = 10^{-6}$ (points
      with error bars).
\end{list}

 
\clearpage
 
\def\cf{$C_F^2$}
\def\cn{$C_FN_C$}
\def\cnf{$C_FN_f$}
\def\nc{\\}
\def\sc{\\[-1mm]}
\small
 
\begin{table}
\caption{~}
\begin{center}
\begin{tabular}{||l|c||c|c|c||c||}
\hline\hline
\multicolumn{6}{||c||}{ NLO Coefficient $B(\chi )$ for $\cos \chi=0$} \\
\hline
$y_{min}$&$method     $&\cf     & \cn    &\cnf    &$sum$\\
\hline
$10^{-6}$ & KL' &$-2.68 \pm 0.68 $   &$78.16 \pm 0.60$&$-25.388 \pm
  0.071 $& $ 50.11 \pm 0.91  $  \\
      & KL   &$-2.57 \pm 0.68 $  &$78.03 \pm 0.60 $ &$ -25.388 \pm
  0.071 $& $ 50.09 \pm 0.91  $  \\
\hline
$10^{-5}$ & KL' &$-2.09 \pm 0.44 $   &$79.22 \pm 0.36$&$-25.159 \pm
  0.055 $& $ 51.96 \pm 0.58  $  \\
      & KL   &$-2.02 \pm 0.44 $  &$79.12 \pm 0.36 $ &$ -25.159 \pm
  0.055 $& $ 51.87 \pm 0.58  $  \\
\hline
$10^{-4}$ & KL'  &$-2.18 \pm 0.26  $ &$80.81 \pm 0.19$ &$-25.116 \pm
  0.041$& $ 53.52 \pm 0.32  $  \\
            &  KL  &$-1.83  \pm 0.26 $ &$80.87 \pm 0.19$ &$-25.116 \pm
  0.041$& $ 53.93 \pm 0.32  $  \\
\hline
$10^{-3}$ & KL'  &$-3.77 \pm 0.13  $ &$81.621\pm 0.092$ &$-24.965 \pm
  0.026$& $ 52.88 \pm 0.16  $  \\
            &  KL  &$-1.61  \pm 0.13 $ &$81.575\pm 0.093$ &$-24.965 \pm
  0.026$& $ 55.00 \pm 0.16  $  \\
\hline
$10^{-2}$ & KL'  &$-11.082\pm 0.054 $ &$80.309\pm 0.063$ &$-24.759 \pm
  0.015$& $ 44.465\pm 0.084 $  \\
            &  KL  &$-2.302 \pm 0.046$ &$79.838\pm 0.065$ &$-24.759 \pm
  0.015$& $ 52.566\pm 0.086 $  \\
\hline\hline
\end{tabular}
\end{center}
\end{table}
 
\begin{table}
\caption{~}
\begin{center}
\begin{tabular}{||l|c||c||}
\hline\hline
\multicolumn{3}{||c||}{NLO Coefficient $B(\chi)$ for $\cos \chi = 0$,
$y_{min} = 10^{-4}$} \\
\hline
$colour factor$&$region$&$B(\cos \chi = 0) $ \\
\hline
 \cf  &sing  &$-410.18\pm 0.16$ \\
      &real  &$ 408.36\pm 0.020$ \\
      &fin   &$-0.3744\pm 0.0019$ \\
      &sum   &$-2.18  \pm 0.26  $ \\
\hline\hline
 \cn  &sing    &$-68.818\pm 0.071$ \\
      &real(13)&$-120.62\pm 0.12 $ \\
      &real(34)&$ 263.80\pm 0.12 $ \\
      &fin(13) &$-0.3023\pm 0.0007$ \\
      &fin(34) &$ 6.706 \pm 0.016 $ \\
      &sum   &$ 80.81 \pm 0.19  $ \\
\hline\hline
 \cnf &sing  &$-44.349\pm 0.018$ \\
      &real  &$ 19.227\pm 0.037$ \\
      &sum   &$-25.116\pm 0.041 $ \\
\hline
 \cf-\cn/2 &$(q\bar{q})_{int}$&$ -0.0851 \pm 0.0004  $ \\
\hline\hline
$sum      $& sum            &$ 53.52  \pm 0.32 $ \\
\hline\hline
\end{tabular}
\end{center}
\end{table}
 
\clearpage
 
\begin{table}
\caption{~}
\begin{center}
\begin{tabular}{||l|c||c|c|c||c||}
\hline\hline
\multicolumn{6}{||c||}{ NLO Coefficient $B(\chi )$ for $\cos \chi=0$} \\
\hline
$y_{min}$&$method     $&\cf     & \cn    &\cnf    &$sum$\\
\hline
$10^{-6}$ & KL' &$-2.67 \pm 0.68 $   &$77.02 \pm 0.60$&$-25.388 \pm
  0.071$& $ 48.98 \pm 0.91  $  \\
      & KL   &$-2.58 \pm 0.68 $  &$76.90 \pm 0.60 $ &$ -25.388 \pm
  0.071$& $ 48.95 \pm 0.91  $  \\
\hline
$10^{-5}$ & KL' &$-2.03 \pm 0.44 $   &$76.26 \pm 0.35$&$-25.159 \pm
  0.055$& $ 49.06 \pm 0.56  $  \\
      & KL   &$-2.04 \pm 0.44 $  &$76.12 \pm 0.35 $ &$ -25.159 \pm
  0.055$& $ 48.92 \pm 0.56  $  \\
\hline
$10^{-4}$ & KL'  &$-1.80 \pm 0.26  $ &$74.39 \pm 0.19$ &$-25.116  \pm
  0.041 $& $ 47.47 \pm 0.32  $  \\
            &  KL  &$-1.89  \pm 0.26 $ &$74.48 \pm 0.19$ &$-25.116  \pm
  0.041 $& $ 47.49 \pm 0.32  $  \\
\hline
$10^{-3}$ & KL'  &$-1.73 \pm 0.13  $ &$69.714 \pm 0.086 $ &$-24.965  \pm
  0.026 $& $ 43.02 \pm 0.15  $  \\
          &  KL  &$-1.86  \pm 0.13 $ &$69.791 \pm 0.087 $ &$-24.965  \pm
  0.026 $& $ 42.97 \pm 0.15  $  \\
\hline
$10^{-2}$ & KL'  &$ -2.610 \pm 0.046  $ &$63.751 \pm 0.041 $ &$-24.759
 \pm 0.015 $& $ 36.380 \pm 0.063  $  \\
          &  KL  &$-3.336  \pm 0.049 $ &$63.887 \pm 0.046 $ &$-24.759
            \pm
  0.015 $& $ 35.791 \pm 0.068  $  \\
\hline\hline
\end{tabular}
\end{center}
\end{table}
 
\begin{table}
\caption{~}
\begin{center}
\begin{tabular}{||l|c||}
\hline\hline
\multicolumn{2}{||c||}{NLO Coefficient $B(\chi)$ for $\cos \chi = 0$} \\
\hline
$calculation  $& $B(\cos \chi = 0) $ \\
\hline
$KN^{(1)}$   &$ 51.25 \pm 2.67$ \\
\hline
 $GS^{(2)}$  &$ 52.39 \pm 0.83$ \\
\hline
 $GS^{(3)}$  &$ 51.15 \pm 0.68$ \\
\hline
 $GS^{(4)}$  &$ 52.29 \pm 2.08 $ \\
\hline
 $KS^{(5)}$  &$ 51.96 \pm 0.58 $ \\
\hline
 $KS^{(6)}$  &$ 51.87 \pm 0.58 $ \\
\hline
 $KS^{(7)}$  &$ 49.06 \pm 0.56 $ \\
\hline
 $KS^{(8)}$  &$ 48.92 \pm 0.56 $ \\
\hline\hline
\end{tabular}
\end{center}
\end{table}
 
\clearpage
 
\setlength{\unitlength}{0.1bp}
\begin{center}
{\bf Fig.\ 1}
\end{center}
\begin{picture}(4320,3067)(0,0)
\put(3628,-49){\makebox(0,0){$\cos \chi$}}
\put(100,2633){%
\makebox(0,0)[b]{\shortstack{$\sin^2\chi B(\chi)$}}%
}
\put(3976,151){\makebox(0,0){1}}
\put(3172,151){\makebox(0,0){0.5}}
\put(2369,151){\makebox(0,0){0}}
\put(1565,151){\makebox(0,0){-0.5}}
\put(761,151){\makebox(0,0){-1}}
\put(540,3016){\makebox(0,0)[r]{1000}}
\put(540,2325){\makebox(0,0)[r]{500}}
\put(540,1634){\makebox(0,0)[r]{0}}
\put(540,942){\makebox(0,0)[r]{-500}}
\put(540,251){\makebox(0,0)[r]{-1000}}
\end{picture}
\begin{center}
{\bf Fig.\ 2}
\end{center}
\begin{picture}(4320,3067)(0,0)
\put(3628,-49){\makebox(0,0){$\cos \chi$}}
\put(100,2633){%
\makebox(0,0)[b]{\shortstack{$\sin^2\chi B(\chi)$}}%
}
\put(3976,151){\makebox(0,0){1}}
\put(3172,151){\makebox(0,0){0.5}}
\put(2369,151){\makebox(0,0){0}}
\put(1565,151){\makebox(0,0){-0.5}}
\put(761,151){\makebox(0,0){-1}}
\put(540,2740){\makebox(0,0)[r]{600}}
\put(540,2187){\makebox(0,0)[r]{400}}
\put(540,1634){\makebox(0,0)[r]{200}}
\put(540,1081){\makebox(0,0)[r]{0}}
\put(540,528){\makebox(0,0)[r]{-200}}
\end{picture}
 
\newpage

\begin{center}
{\bf Fig.\ 3}
\end{center}
\begin{picture}(4320,3067)(0,0)
\put(3628,-49){\makebox(0,0){$\cos \chi$}}
\put(100,2633){%
\makebox(0,0)[b]{\shortstack{$\sin^2\chi B(\chi)$}}%
}
\put(3976,151){\makebox(0,0){1}}
\put(3172,151){\makebox(0,0){0.5}}
\put(2369,151){\makebox(0,0){0}}
\put(1565,151){\makebox(0,0){-0.5}}
\put(761,151){\makebox(0,0){-1}}
\put(540,3016){\makebox(0,0)[r]{40}}
\put(540,2621){\makebox(0,0)[r]{20}}
\put(540,2226){\makebox(0,0)[r]{0}}
\put(540,1831){\makebox(0,0)[r]{-20}}
\put(540,1436){\makebox(0,0)[r]{-40}}
\put(540,1041){\makebox(0,0)[r]{-60}}
\put(540,646){\makebox(0,0)[r]{-80}}
\put(540,251){\makebox(0,0)[r]{-100}}
\end{picture}
\begin{center}
{\bf Fig.\ 4}
\end{center}
\begin{picture}(4320,3067)(0,0)
\put(3628,-49){\makebox(0,0){$\cos \chi$}}
\put(100,2633){%
\makebox(0,0)[b]{\shortstack{$\sin^2\chi B(\chi)$}}%
}
\put(3976,151){\makebox(0,0){1}}
\put(3172,151){\makebox(0,0){0.5}}
\put(2369,151){\makebox(0,0){0}}
\put(1565,151){\makebox(0,0){-0.5}}
\put(761,151){\makebox(0,0){-1}}
\put(540,2670){\makebox(0,0)[r]{200}}
\put(540,2238){\makebox(0,0)[r]{150}}
\put(540,1806){\makebox(0,0)[r]{100}}
\put(540,1374){\makebox(0,0)[r]{50}}
\put(540,942){\makebox(0,0)[r]{0}}
\put(540,510){\makebox(0,0)[r]{-50}}
\end{picture}

\newpage 

\begin{center}
{\bf Fig.\ 5}
\end{center}
\begin{picture}(4320,3067)(0,0)
\put(3778,-49){\makebox(0,0){$y_{min}$}}
\put(160,2633){%
\makebox(0,0)[b]{\shortstack{$B(\pi/2)$}}%
}
\put(4137,151){\makebox(0,0){0.1}}
\put(3548,151){\makebox(0,0){0.01}}
\put(2958,151){\makebox(0,0){0.001}}
\put(2369,151){\makebox(0,0){0.0001}}
\put(1779,151){\makebox(0,0){1e-05}}
\put(1190,151){\makebox(0,0){1e-06}}
\put(600,151){\makebox(0,0){1e-07}}
\put(540,3016){\makebox(0,0)[r]{20}}
\put(540,2740){\makebox(0,0)[r]{18}}
\put(540,2463){\makebox(0,0)[r]{16}}
\put(540,2187){\makebox(0,0)[r]{14}}
\put(540,1910){\makebox(0,0)[r]{12}}
\put(540,1634){\makebox(0,0)[r]{10}}
\put(540,1357){\makebox(0,0)[r]{8}}
\put(540,1081){\makebox(0,0)[r]{6}}
\put(540,804){\makebox(0,0)[r]{4}}
\put(540,528){\makebox(0,0)[r]{2}}
\put(540,251){\makebox(0,0)[r]{0}}
\end{picture}
 
\newpage

\begin{center}
{\bf Fig.\ 6}
\end{center}
\begin{picture}(4320,5227)(0,0)
\put(5728,-6902){\makebox(0,0)[r]{'ycut4.gdt'}}
\put(5728,-6802){\makebox(0,0)[r]{'ycut3.gdt'}}
\put(5728,-6702){\makebox(0,0)[r]{'ycut2.gdt'}}
\put(5728,-6602){\makebox(0,0)[r]{'ycut1.gdt'}}
\put(3572,867){\makebox(0,0)[l]{KL'}}
\put(3572,497){\makebox(0,0)[l]{KL}}
\put(3572,2508){\makebox(0,0)[l]{KL'}}
\put(3572,4150){\makebox(0,0)[l]{KL}}
\put(3868,-49){\makebox(0,0){$y_{min}$}}
\put(160,4013){%
\makebox(0,0)[b]{\shortstack{$B(\pi/2)$}}%
}
\put(4137,151){\makebox(0,0){0.1}}
\put(3548,151){\makebox(0,0){0.01}}
\put(2958,151){\makebox(0,0){0.001}}
\put(2369,151){\makebox(0,0){0.0001}}
\put(1779,151){\makebox(0,0){1e-05}}
\put(1190,151){\makebox(0,0){1e-06}}
\put(600,151){\makebox(0,0){1e-07}}
\put(540,4560){\makebox(0,0)[r]{55}}
\put(540,3534){\makebox(0,0)[r]{50}}
\put(540,2508){\makebox(0,0)[r]{45}}
\put(540,1482){\makebox(0,0)[r]{40}}
\put(540,456){\makebox(0,0)[r]{35}}
\end{picture}

\newpage

\begin{center}
{\bf Fig.\ 7}
\end{center}
\begin{picture}(4320,3067)(0,0)
\put(82755,-2326){\makebox(0,0)[r]{'eez6b.gdt'}}
\put(82755,-2226){\makebox(0,0)[r]{'eez4b.gdt'}}
\put(3568,-49){\makebox(0,0){$\cos \chi$}}
\put(160,2533){%
\makebox(0,0)[b]{\shortstack{$B(\chi)$}}%
}
\put(3976,151){\makebox(0,0){1}}
\put(3172,151){\makebox(0,0){0.5}}
\put(2369,151){\makebox(0,0){0}}
\put(1565,151){\makebox(0,0){-0.5}}
\put(761,151){\makebox(0,0){-1}}
\put(540,3016){\makebox(0,0)[r]{1000}}
\put(540,1389){\makebox(0,0)[r]{100}}
\end{picture}
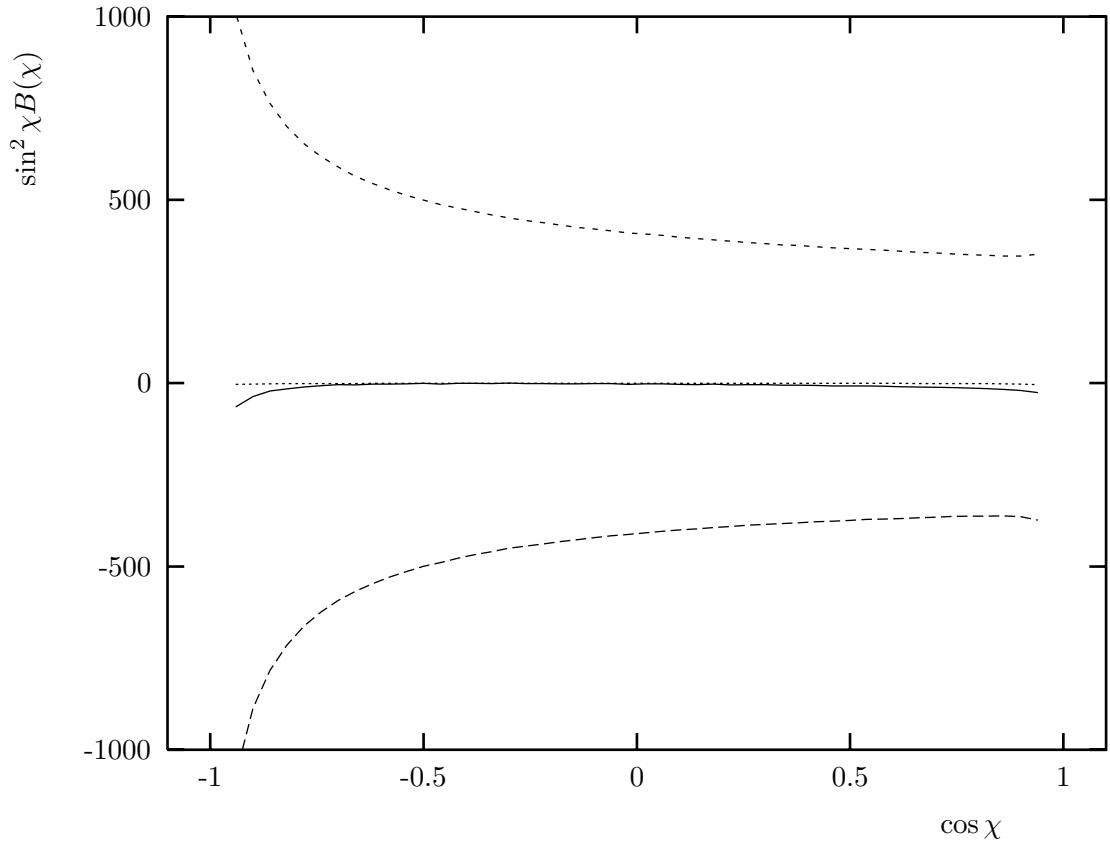
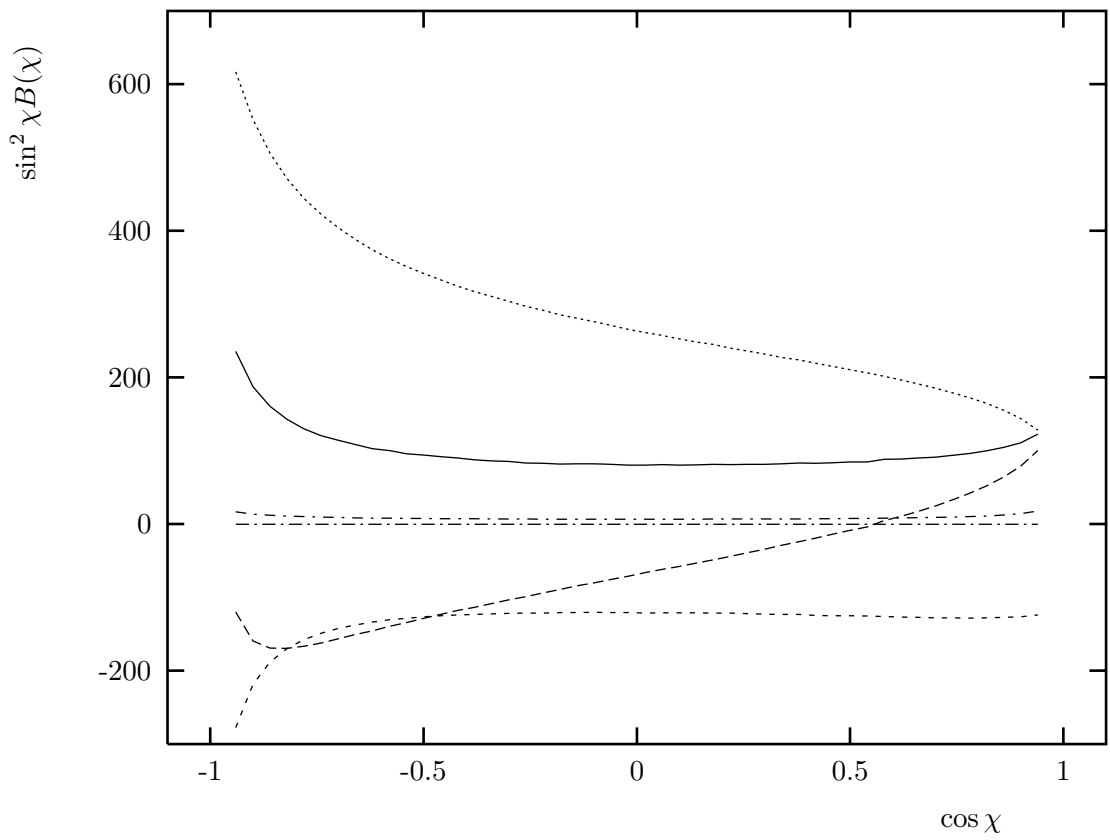
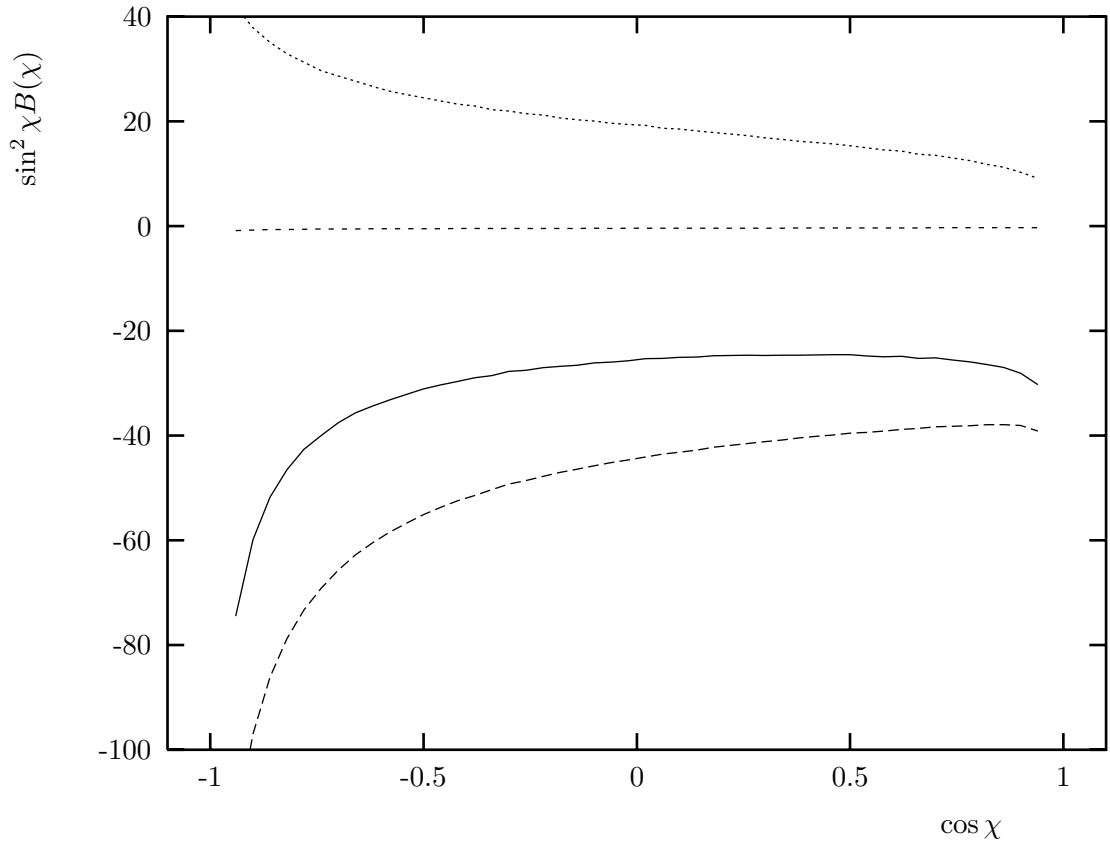
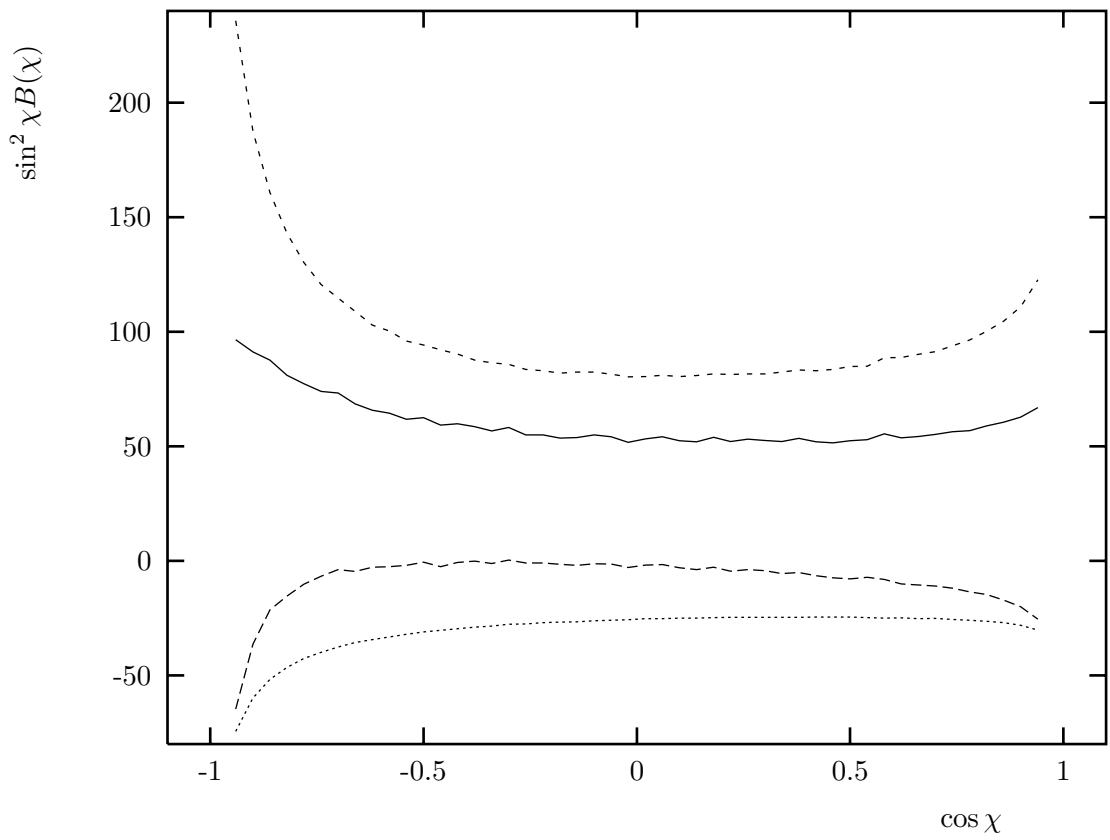
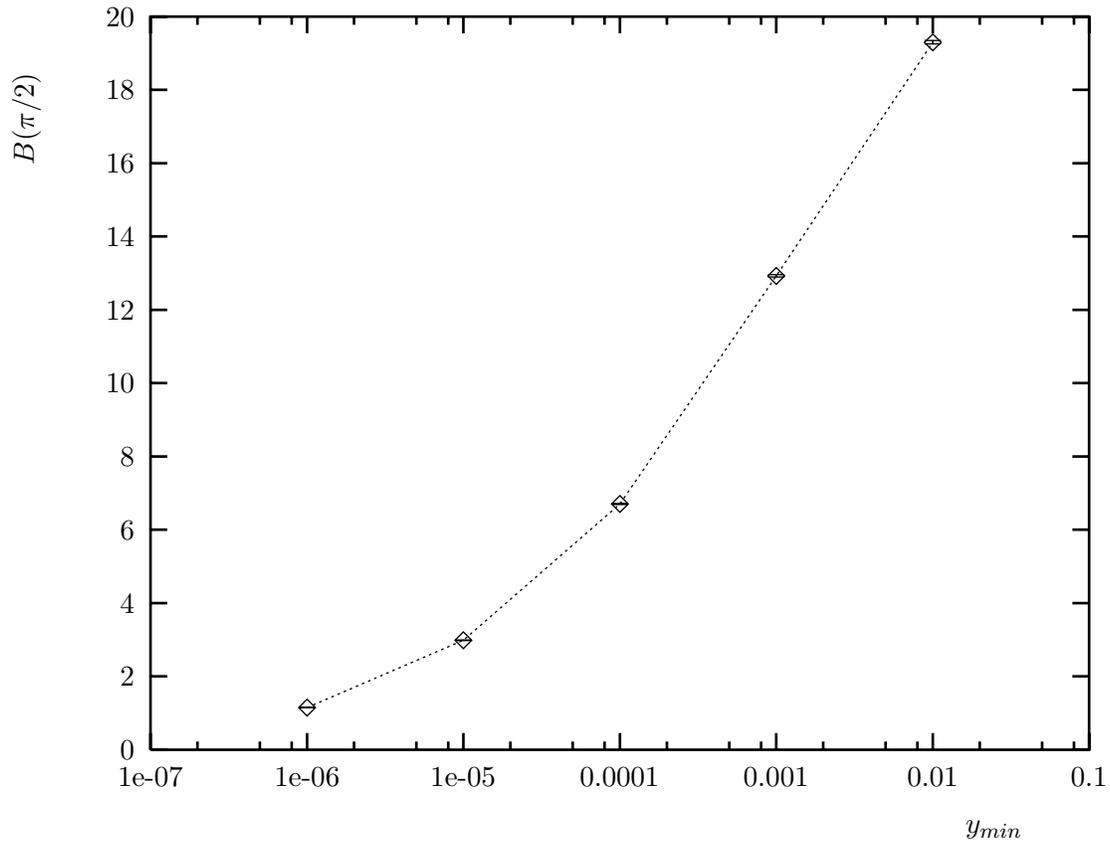
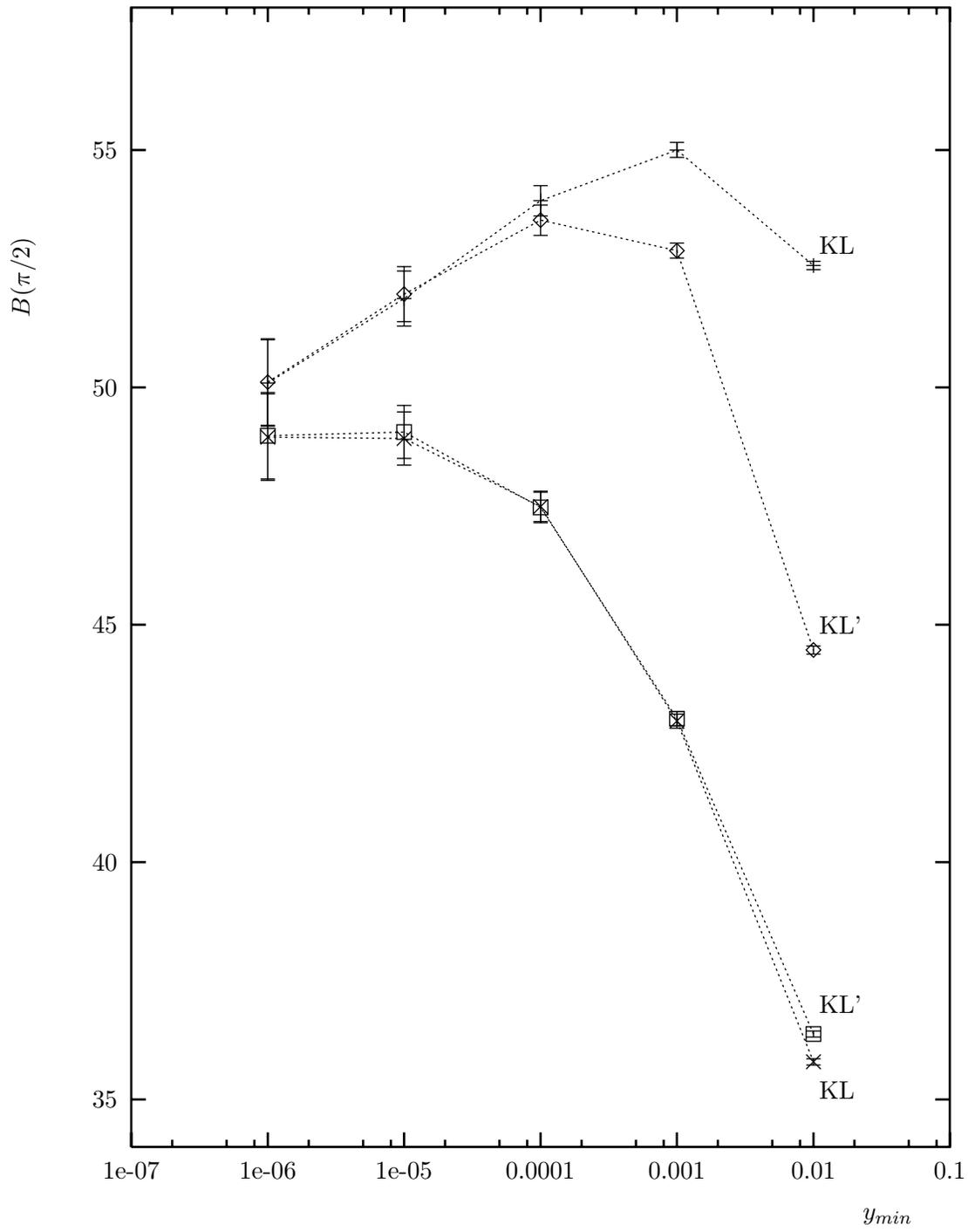
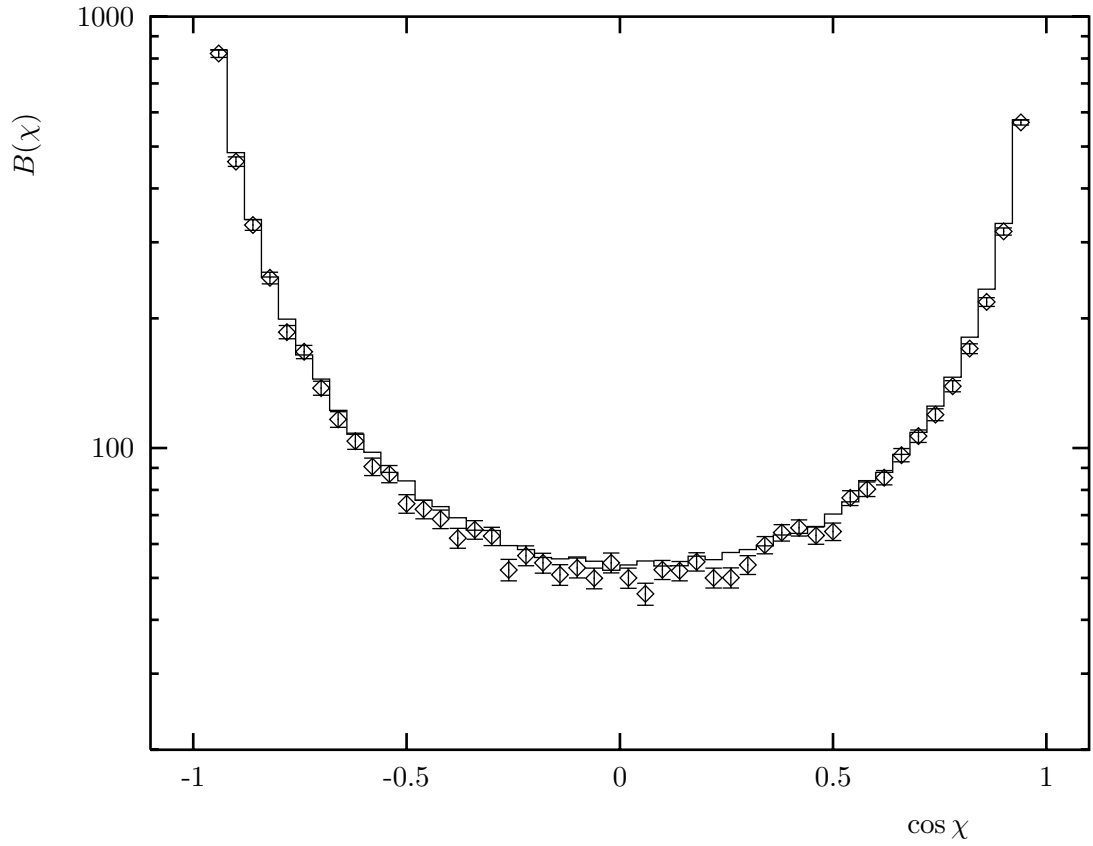
\end{document}